# Content-based and Algorithmic Classifications of Journals:

# Perspectives on the Dynamics of Scientific Communication and Indexer Effects

Ismael Rafols[1] and Loet Leydesdorff [2]

**Abstract**

The aggregated journal-journal citation matrix—based on the *Journal Citation Reports* (JCR) of the *Science Citation Index*—can be decomposed by indexers and/or algorithmically. In this study, we test the results of two recently available algorithms for the decomposition of large matrices against two content-based classifications of journals: the ISI Subject Categories and the field/subfield classification of Glänzel & Schubert (2003). The content-based schemes allow for the attribution of more than a single category to a journal, whereas the algorithms maximize the ratio of within-category citations over between-category citations in the aggregated category-category citation matrix. By adding categories, indexers generate between-category citations, which may enrich the database, for example, in the case of inter-disciplinary developments. The consequent indexer effects are significant in sparse areas of the matrix more than in denser ones. Algorithmic decompositions, on the other hand, are more heavily skewed towards a relatively small number of categories, while this is deliberately counter-acted upon in the case of content-based classifications. Because of the indexer effects, science policy studies and the sociology of science should be careful when using content-based classifications, which are made for bibliographic disclosure, and not for the purpose of analyzing latent structures in scientific communications. Despite the large differences among them, the four classification schemes enable us to generate surprisingly similar maps of science at the global level. Erroneous classifications are cancelled as noise at the aggregate level, but may disturb the evaluation locally.

---

[1] SPRU –Science and Technology Policy Research, University of Sussex, Brighton, BN1 9QE, England; i.rafols@sussex.ac.uk ; http://www.sussex.ac.uk/spru/irafols .
[2] Amsterdam School of Communication Research (ASCoR), University of Amsterdam, Kloveniersburg 48, 1012 CX Amsterdam, The Netherlands; loet@leydesdorff.net ; http://www.leydesdorff.net .

**Introduction**

In the summer of 2008, Vincent Blondel and his colleagues published a new algorithm for the decomposition of large networks at *arXiv* (Blondel *et al*., 2008). This algorithm was quickly adopted in a follow-up study by Wallace *et al*. (2008) for the study of research communities in terms of citation networks. Others followed as the algorithm was made publicly available at http://findcommunities.googlepages.com (e.g., Lambiotte & Panzarasa, 2008). Another algorithm for the same purpose was recently made available by Martin Rosvall at http://www.tp.umu.se/~rosvall/code.html. Rosvall developed this program in the framework of a study (with Carol Bergstrom) of the aggregated journal-journal citation matrix using 6,128 journals included in the *Journal Citation Reports* (JCR) of 2004 (Rosvall & Bergstrom, 2008).

At that time, we (Leydesdorff & Rafols, forthcoming) were deeply involved in testing the ISI Subject Categories of these same journals in terms of their disciplinary organization. Using the JCR of the *Science Citation Index* (SCI) we found 14 major components using 172 subject categories, and 6,164 journals in 2006. Given our analytical objectives, and the well-known differences in citation behaviour within the social sciences (Bensman, 2008), we decided to set aside the study of the (220 – 175 =) 45 subject categories in the social sciences for a future study.³

Our findings using the *Science Citation Index* indicated that the ISI Subject Categories can be used for statistical mapping purposes at the global level despite being imprecise in terms of the detailed attribution of journals to the categories. Might it be that the two newly available algorithms would provide us with more reliable classifications of the journals? Rosvall & Bergstrom (2008), for example, claimed to find 88 components using the same data for 2004. What effect would Blondel *et al.*'s (2008) new unfolding algorithm have on the JCR data?

In addition to these three decompositions—that is, the ISI Subject Classifications, the results of using Rosvall & Bergstrom's (2008) maps of random walks, and Blondel *et al.*'s (2008) unfolding algorithm—a fourth classification system of journals was proposed by Glänzel & Schubert (2003) and increasingly used for evaluation purposes by the Steungroep Onderwijs & Onderzoek Indicatoren (SOOI) in Leuven, Belgium. These authors originally proposed 12 fields and 60 subfields for the *Science Citation Index,* and 3 fields and 7 subfields for the *Social Science Citation Index* and the *Arts and Humanities Citation Index*. Later, one more subfield entitled "multidisciplinary sciences" was added.

Wolfgang Glänzel very kindly provided us with a copy of this classification of journals for the purpose of this research (Glänzel, *personal communication,* October 11, 2008).

---

³ Three (of the 175) subject categories in the *Science Citation Index* are no longer actively used for the indexing.

Based on our replication of Rosvall & Bergstrom's analysis using 2006 data, and on our collaboration with Renaud Lambiotte—one of the authors of Blondel *et al.* (2008)—we are able to compare below the four classification schemes using the same dataset of the combined JCRs for the *Social Science Citation Index* and the *Science Citation Index* 2006.

**The importance of journal classifications**

Garfield's (1971) "law of concentration" can be considered as a corollary of Bradford's (1934) "law of scattering." Whereas Bradford's law formulated that "articles of interest to a specialist must occur not only in the periodicals specialising in his subject," Garfield stated that the tails of journals' citation distributions can be expected to encompass the cores of other sets (Garfield, 1972). Thus, because research topics are on the one hand thinly spread outside the core group and on the other hand the core groups are interwoven, one cannot expect that the aggregated journal-journal citation matrix matches one-to-one with substantive definitions of categories or that it can be decomposed in a single and unique way in relation to scientific specialties. The choice of an appropriate journal set can be considered as a local optimization problem (Leydesdorff, 2006).

Citation relations among journals are dense in discipline-specific clusters and are otherwise very sparse, to the extent of being virtually non-existent (Leydesdorff & Cozzens, 2003).[4] Although not completely decomposable, the matrix can be considered as heavily structured and nearly decomposable. The grand matrix of aggregated journal-journal citations is so heavily structured that the mappings and analyses in terms of citation distributions have been amazingly robust despite differences in methodologies (e.g., Leydesdorff, 1987 and 2007; Tijssen *et al*., 1987; Boyack *et al*., 2005; Moya-Anegón *et al.*, 2007; Klavans & Boyack, forthcoming).

---

[4] In 2006, the database contained only 1,315,143 of the 57,927,321 (= $7611^2$) possible relations between two journals, or in other words a density of 2.27%.

A decomposable matrix is a square matrix such that a rearrangement of rows and columns leaves a set of square sub-matrices on the principal diagonal and zeros everywhere else. In the case of a nearly-decomposable matrix some zeros are replaced by relatively small nonzero numbers (Simon & Ando, 1961; Ando & Fisher, 1963). Near-decomposability is a general property of complex and evolving systems (Simon, 1973 and 2002). The next-order units represented by the square sub-matrices—and representing in this case disciplines or specialties—are reproduced in relatively stable sets (of journals) which may change over time. The sets of journals are functional subsystems that show a high density in terms of relations within the centre (that is, core journals), but are more open to change in relations at the margins. The organization among the subsystems can also change.

The decomposition into nearly-decomposable matrices has no analytical solution. However, algorithms can provide heuristic decompositions when there is no single unique correct answer. Newman (2006a; 2006b) proposed using *modularity* for the decomposition of nearly decomposable matrices since modularity can be maximized as an objective function. Blondel *et al.* (2008) used this function for relocating units iteratively in neighbouring clusters. Each decomposition can then be considered in terms of whether it increases the modularity. Analogously, Rosvall & Bergstrom (2008) maximized the probabilistic entropy between clusters by estimating the fraction of time during which every node is visited in a random walk (cf. Theil, 1972; Leydesdorff, 1991).

**Methods and data**

The data were harvested from the CD-Rom version of the *Journal Citation Reports* of the *Science Citation Index* and *Social Science Citation Index* 2006, and then combined. In 2006, the *Journal Citation Report* for the *Science Citation Index* contained 6,164 journals, and the *Social Science Citation Index* 1,768 journals. We corrected for the overlap of 321 journals which are contained in both databases. The resulting set of 7,611 journals and their citation relations otherwise precisely corresponds to the online version of the *Journal Citation Reports.* This large data matrix of 7,611 times 7,611 citing and cited journals was stored conveniently as a Pajek (.net) file and used for further processing.

The 7,611 journals are attributed by the ISI with 11,856 subject classifiers. This is 1.56 ($\pm$ 0.76) classifiers per journal. The 220 ISI Subject Categories are assigned by ISI staff on the basis of a number of criteria including the journal's title and its citation patterns (McVeigh, *personal communication*, 9 March 2006; Bensman & Leydesdorff, forthcoming). Pudovkin and Garfield (2002) state that the ISI assigns journals to categories by "subjective, heuristic methods" (p. 1113):

> …This method is "heuristic" in that the categories have been developed by manual methods started over 40 years ago. Once the categories were established, new journals were assigned one at a time. Each decision was based upon a visual examination of all relevant citation data. As categories grew, subdivisions were established. Among other tools used to make individual journal assignments, the Hayne-Coulson algorithm is used. The algorithm has never been published. It treats any designated group of journals as one macrojournal and produces a combined printout of cited and citing journal data. (Pudovkin and Garfield, 2002, at p. 1113n.)[5]

According to the evaluation of Pudovkin and Garfield (2002), in many fields these categories are sufficient, but the authors added that "in many areas of research these 'classifications' are crude and do not permit the user to quickly learn which journals are most closely related" (p. 1113). Boyack *et al*. (2005) estimated that the attribution is correct in approximately 50% of cases across the file (Boyack, *personal communication,* 14 September 2008). Leydesdorff & Rafols (forthcoming) found that the ISI Subject Categories can be used for statistical purposes—the factor analysis for example can remove the noise—but not for the detailed evaluation. In the case of interdisciplinary fields, problems of imprecise or potentially erroneous classifications can be expected.

For the purpose of developing a new classification scheme of scientific journals contained in the *Science Citation Indices*, Glänzel & Schubert (2003) used three successive steps for their attribution. The authors iteratively distinguished sets *cognitively* on the basis of expert judgements, *pragmatically* in order to retain multiple assignments within reasonable limits, and *scientometrically* using unambiguous core journals for the classification. The scheme of 15 fields and 68 subfields is used extensively for research evaluations by the Steunpunt Onderwijs & Onderzoek Indicatoren (SOOI), a research unit at the Catholic University in Leuven, Belgium, headed by Glänzel.

---

[5] Pudovkin & Fuseler (1995, p. 228) further specified the Hayne-Coulson algorithm as follows: "The number of citations each journal receives from different specialty core journals is obtained annually by a computer routine (Hayne-Coulson) that is used to create the JCR database."

The SOOI categories cover 8,985 journals. Using the full titles of the journals, 7,485 could be matched with the 7,611 journals under study in the *JCR* data for 2006 (which is 98.3%). These journals are attributed 10,840 classifiers at the subfield level. This is 1.45 ($\pm$ 0.66) categories per journal. One category ("Philosophy and Religion") is missing because the *Arts & Humanities Citation Index* is not included in our data. Thus, we pursued the analysis with the 67 SOOI categories.

Like the ISI Subject Categories, the SOOI categories are kept the same over time in order to make longitudinal comparisons possible. In other words, one can also consider the classification by Glänzel & Schubert (2003) as an update of the old (1974) ISI classification. Neither classification, however, can be reproduced by an outsider. Furthermore, they are not based on "literary warrant" like the classification of the Library of Congress (Chan, 1999). The LC has a policy of continuous revision in order to take current literary warrant into account, so that new areas are developed and obsolete elements are removed or revised (Leydesdorff & Bensman, 2006: 1473). The ISI or SOOI categories, however, are changed in terms of respective coverage, but cannot themselves be revised from the perspective of hindsight (Leydesdorff, 2002).

Unlike the content-based classifications, the two algorithms under study provide unique assignments of journals to groups. Using Rosvall & Bergstrom's (2008) algorithm with 2006 data, we obtained findings similar to those of these authors on August 11, 2008. Like the original authors using 6,128 journals in 2004, we found 88 clusters using 7,611 journals in 2006. We used this attribution for the purposes of further analysis.[6] Lambiotte, one of the coauthors of Blondel *et al*. (2008), was so kind as to input the data into the unfolding algorithm and found the following results: 114 communities with a modularity value of 0.527708 and 14 communities with a modularity value of 0.60345. We use the 114 communities for the purposes of this comparison. These categories refer to 7,607 (= 7611 - 4) journals because four of the journals in the file were isolates.[7]

**Results**

In order to understand the differences and similarities among the four classifications investigated, we proceed with the following analyses. First, the statistical properties of each of the decompositions are investigated. Secondly, we look at the degree of agreement between the decompositions. Thirdly, the visual maps of science obtained using the diverse classifications are compared. In the conclusions and discussion sections, we discuss the main differences between the four classification schemes and the implications for information retrieval and the sociology of science.

---

[6] In response to our exchange, Rosvall uploaded an improved algorithm on August 13, 2008, which provides different numbers of clusters depending on a random seed number. This algorithm is no longer deterministic.

[7] These four journals are: *Mer-Marine Engineers Review, Man in India, Nursing History Review,* and *Russian Politics and Law*.

*Statistical properties of the decompositions*

Let us begin by providing the descriptive statistics of the different classifications used in this study. The number of categories in each decomposition is of the same order of magnitude of hundred—with a range from 67 (SOOI) to 220 (ISI) as the extremes—and therefore one can expect that the results will be roughly comparable. The number of journals per category is log-normally distributed in each of the four classifications. In other words, they all have a relatively small number of categories with a large number of journals, and many categories with only a few journals. However, as shown in Figure 1, the classifications based on the Random Walk and Unfolding algorithms are more skewed than the content-based classifications.

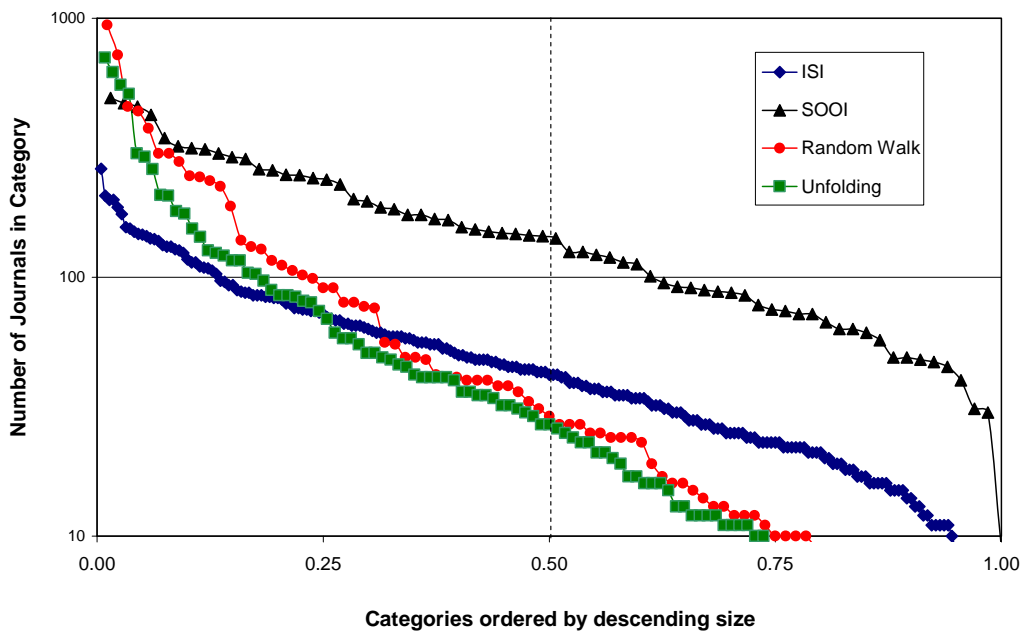

**Figure 1**. Number of journals in each category ranked according to size. The horizontal axis is normalized so that the category with the highest number of journals appears in 0 and the one with the lowest value in 1.

Whereas the top 10 categories on the basis of a Random Walk comprise 57% of the journals (50% for Unfolding), they cover only 15% in the ISI decomposition and 31% for the SOOI classification. In the case of skewed distributions, the characteristic number of journals per category can best be expressed by the median: the median is below 30 in the case of the Random Walk or Unfolding, compared to 42 journals for the ISI classification, and 141 for the SOOI classification (Table 1).

| *7,611 journals* | ISI | SOOI | *Random Walk* | *Unfolding* |
|---|---|---|---|---|
| Number of Categories | 220 | 67 | 88 | 114 |
| Journals/Categ. | | | | |
| Mean | 53 | 161 | 86 | 66 |
| Median | 42 | 141 | 28 | 26 |
| Standard deviation | 43 | 113 | 151 | 118 |
| Citations Within/Between Categories | | | | |
| Within Category citations | 13,286,544 | 14,194,033 | 15,458,430 | 13,630,264 |
| Between Category citations | 41,163,842 | 33,013,188 | 7,412,325 | 9,240,491 |
| Grand-sum of citation matrix | 54,450,386 | 47,207,221 | 22,870,755 | 22,870,755 |
| Within/Between Categories Ratio | 3.10 | 2.33 | 0.48 | 0.68 |

**Table 1.** Statistical properties of the number of journals per category, and the within/between category citations for each decomposition.

Another way to analyze the overlaps among categories for a given decomposition is to compare the number of aggregated citations between categories (off-diagonal) to the citations within categories (on the diagonal). As presented in the last rows of Table 1, the total numbers of citations in the aggregated matrices based on the ISI or SOOI classifications are much higher because the same citation can be attributed to two or three categories. Thus, whereas Random Walk and Unfolding lead to matrices with most citations within categories (on the diagonal), matrices based on ISI and SOOI classifications lead to matrices with most citations between categories (off-diagonal).

Finally, in order to measure how similar the categories in the four decompositions are to each other, we computed the Cosine Similarities in the citation patterns between each pair of citing categories in the four aggregated category-category matrices (Salton & McGill, 1983; Ahlgren *et al*., 2003). These Cosine Similarities provide an indicator between zero and one of the degree to which two categories in a decomposition cover the same set of journals. For example, in the ISI decomposition, the new *Nanoscience and Nanotechnology* category has a similarity of cosine = 0.003 with *Women's Studies* (hence they are completely distinct), but a cosine = 0.50 similarity with *Electrochemistry* (suggesting important overlap), and a cosine = 0.96 with *Materials Science, Interdisciplinary*. This high value raises the question of whether these two categories, as defined by the ISI, should be distinguished from one another.

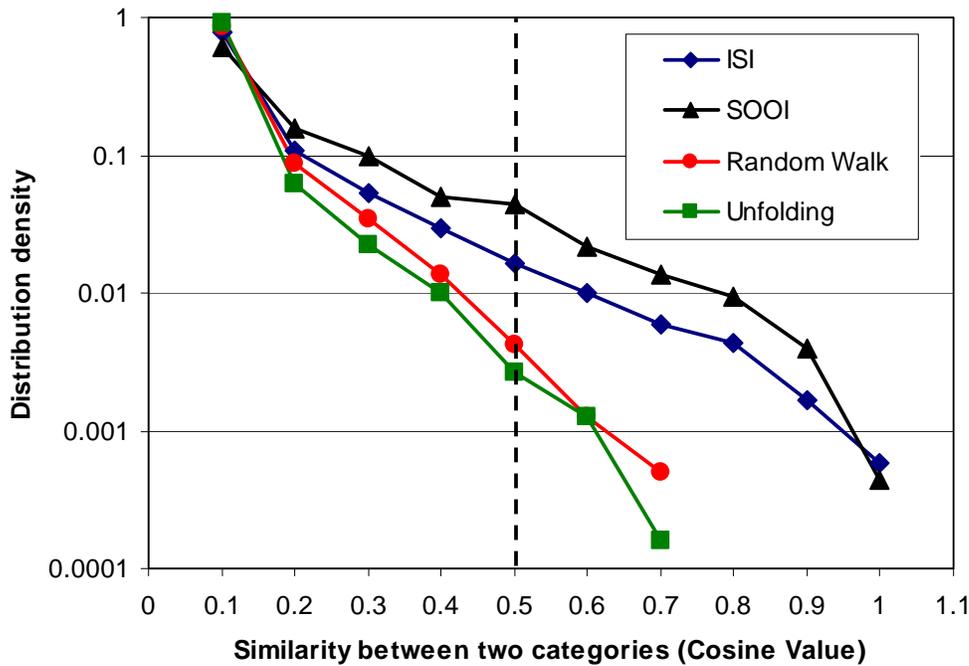

**Figure 2**. Distribution density of cosine similarities between categories. ISI and SOOI's similarity values are much higher than those of a Random Walk or the Unfolding algorithm.

| Cosine Similarity between Categories | ISI | SOOI | Random Walk | Unfolding |
|---|---:|---:|---:|---:|
| Number of Categories | 220 | 67 | 88 | 114 |
| Mean | 0.076 | 0.136 | 0.044 | 0.033 |
| Median | 0.020 | 0.066 | 0.009 | 0.007 |
| Stand. Dev. | 0.129 | 0.165 | 0.077 | 0.065 |

**Table 2.** Statistical properties of the cosine similarities between categories and the within/between category citations for each decomposition.

The statistical analyses of the cosine similarities are presented in Table 2 and Figure 2, which show the density distribution of the similarities between categories, i.e., the proportion of similarities that fall within a certain value bin (e.g. in the case of Unfolding, the proportion of similarities between 0.3 and 0.4 is 0.01, or 1% of the similarities). We find again that all the distributions are highly skewed and that the Random Walk and Unfolding algorithms exhibit a much lower median similarity value among categories. The lower medians indicate that the algorithmic decompositions produce a much "cleaner" cut between categories than the content-based classifications. As shown in Figure 2 the density dissimilarity values above 0.5 is below the 0.01 proportion (or 1%) in the algorithmic decompositions (0.6% in Random Walk and 0.4% in Unfolding), whereas it is ten times higher using the content-based classifications (3.9% for ISI and 9.3% in SOOI).

In conclusion, the analysis of the statistical properties of the different classifications teaches us that the Random Walk and the Unfolding algorithms produce much more skewed distributions in terms of the number of journals per category, but these constructs are more specific than the content-based classification of the ISI and SOOI.

*Degree of agreement between different decompositions*

Let us now turn to the relations among the four decompositions. First, we tested whether the distributions are independent. To this end, the four classifications were cross-tabled, and the cross-tables tested for the zero-hypothesis of mutual independence by using chi-square (with Yule-correction because of the low expected values in many cells). This zero-hypothesis was rejected in all comparisons at the level of $p < 0.001$. Not surprisingly, the four classification schemes are correlated.

Secondly, we focused on the diagonal values in the cross-tables. Appendix A provides an example of how the cross-tables are analyzed. If two classification schemes have a perfect match between them—which cannot be expected because of the different numbers of categories—the results of cross-tabling would yield a matrix with values that can be sorted on the main diagonal; all off-diagonal values would then be zero. Therefore, the proportion of journals on the diagonal of the cross-tables provides us with a measure of the accuracy of the fit between matching categories from different decompositions.

| *% journals given this decomposition...* | *...in top matching category of this decomposition* | | | |
|---|---|---|---|---|
| | ISI | SOOI | Random Walk | Unfolding |
| ISI | - | 98.2 | 66.2 | 60.1 |
| SOOI | 58.4 | - | 51.5 | 47.0 |
| Random Walk | 43.7 | 56.0 | - | **71.3** |
| Unfolding | 48.4 | 59.3 | **85.7** | - |

**Table 3**. Percentage of journals in one decomposition belonging to the matching category in another decomposition.

Table 3 shows the level of matching. For example, given the categorization by the Random Walk algorithm (third row), only 43.7% of the journals are located in the main matching category of the ISI classification (first column). Since journals in the classification schemes of the ISI and SOOI are assigned to more than a single category, these percentages provide an over-estimation of the agreement because a larger chance of a match is generated by the multiple assignments. Furthermore, one can expect the agreement to be higher when categories from a finer decomposition are used as a predictor for a coarser one.

For example, when the 220 ISI Subject Categories are used as predictors for the 67 SOOI subfields, the ISI-to-SOOI matching is 98.2%, whereas the SOOI-to-ISI matching is only 58.0% despite the favorable effect of multiple assignments in both schemes. Not surprisingly, the agreement among diagonal elements is highest (71.3% and 85.7%; bold-faced in Table 3) when comparing the results of the Random Walk and the Unfolding algorithm because these are both based on the maximization of an objective function.

Table 3 indicates that the concordance between the different classifications is modest: in general in the 40-60% range, which is in agreement with Boyack's estimation of 50% correct classifications for the ISI categories (Boyack, *personal communication,* 14 September 2008). However, when we expand the diagonal from the single most matching category to the top three matching categories, we obtain agreement in the order of 70-85%, that is, a 15-25% increase in comparison to the figures in Table 3. For example, the ISI-to-Random Walk agreement is in this case 80.5%, ISI-to-Unfolding is 84.3%, SOOI-to-Random Walk is 78.4%, and SOOI-to-Unfolding is 74.0%.

In summary, although the correspondences among the main categories are sometimes as low as 50% of the journals, most of the mismatched journals appear to fall in areas within the close vicinity of the main categories. In other words, it seems that the various decompositions are roughly consistent, but imprecise. We believe that this broad-range agreement, in spite of local inaccuracies, explains the unexpected similarities among the science maps that we present below.

*Comparison between science maps*

Thus far we have investigated, first, the fit within each decomposition of the journal-to-journal citation matrix and, second, the fit between different decompositions in terms of matching categories. Let us now proceed to explore the agreement between different decompositions in terms of how the respective categories relate to each other in a network. For this purpose, we generated maps of science using the four different decompositions. Since the similarities used to make these maps depend on cross-category citations, a structural agreement in the map would indicate concordance in the underlying relations among categories. In other words, the resemblance among these maps provides us with an assessment of the agreement in the off-diagonal elements in the aggregated citation matrices among categories.

Maps of science for each decomposition were generated from the aggregated category-category citation matrices using the cosine as similarity measure. The similarity matrices were visualized with Pajek (Batagelj and Mrvar, 1998) using Kamada & Kawai's (1989) algorithm. The method is kept similar to that used in Leydesdorff and Rafols (forthcoming). However, the threshold value of similarity for edge visualization is pragmatically set at cosine > 0.01 for the algorithmic decompositions and cosine > 0.2 for the content-based decompositions in order to enhance the readability of the maps without affecting the representation of the structures in the data.

Since maps with more than 30-40 labels are difficult to read, the following simplified representations were prepared. For the ISI decomposition, the 220 categories (Figure 3) were clustered into 18 macro-categories (Figure 4) obtained from the factor analysis (cf. Leydesdorff and Rafols, forthcoming). For the SOOI classification, we used the 16 field categories created by the indexers themselves as a grouping of the 67 subfield categories (Figures 6 and 5, respectively). Taking advantage of the concentration of journals in a few categories, in the case of Random Walk and Unfolding only the top 30 and 35 categories were used, respectively. These top-segments comprise 85.9% and 81.4% of all journals, respectively.[8] The labels for the Random Walk and Unfolding classifications (in Figures 7 and 8) were assigned on the basis of on an inspection of the journal titles in the category, but as far as possible in accordance with the labels of the ISI Subject Categories.[9]

Each map shows the categories of each decomposition, their relative position according to their main inter-relations and the size of each category (the area of a node/discipline is proportional to its number of outbound citations, that is, "total citing"). The overall structure of science obtained is astonishingly robust, given the major differences among the four decompositions shown in Table 3. These results do not solve the problems of decomposition, but they support the observation that 50 percent correspondence in the classification is enough for obtaining similar maps using the main dimensions of the network (Leydesdorff and Rafols, forthcoming).

---

[8] The choice of 30 and 35 categories is arbitrary. By using this segment, we capture most of the journals (>80%), and yet the results remain readable.
[9] In about 10% of the cases, the ISI attributions suggested minor corrections to the labels. For example, the field of 'Clinical Neurology' in the Random Walk classification had originally been categorized by us as 'Neurosciences' after inspection of the titles of the journals involved, but was relabeled after finding a 59% correspondence with the ISI Subject Category of 'Clinical Neurology'.

Although the objective of this study is to investigate the differences among decomposition, let us briefly outline a number of findings relevant to the current discussion on the robustness and convergence ("consensus") of science maps in the bibliometrics community (Anegón *et al*., 2007; Klavans and Boyack, forthcoming). All these maps display two main poles: a very large pole in the biomedical sciences, and a second pole in the physical sciences and engineering. These two poles are connected via three bridging areas: chemistry, a geosciences-environment-ecology group, and the computer sciences. The social sciences are somewhat detached, linked via the behavioral sciences/neuroscience to the biomedical pole, and via the computer sciences and mathematics to the physics/engineering pole.[10]

Another way to look at these maps is to use a circular ordering of disciplines as a first approximation. This perspective was recently proposed as overarching by Klavans and Boyack (forthcoming). The circle of disciplines obtained in all four representations is consistent with their circular Consensus Map. Moving in the clockwise direction in Figures 3 to 8, one can identify the following categories: Biomedical Sciences, Biology and Environment, Geosciences and Chemistry, Physics and Engineering, Mathematics and Computer Sciences, Social Sciences, Psychology, Neurosciences and Health Services, Medical Specialties and Infectious Diseases, to Biomedical Sciences again. Note that the use of the word "and" above does not designate a joint category, but two distinct categories that occupy parallel positions in the clockwise reading of the maps.

---

[10] Since we use the JRC data, our data do not contain journals from the *Arts and Humanities Citation Index*.

Why is the overall structure of these science maps so robust despite the large differences in the decompositions? As noted above, although categories of different decompositions do not always match with one another, most "misplaced" journals are assigned into closely neighbouring categories. Therefore, the error in terms of categories is not large and is also unsystematic. The noise-to-signal ratio becomes much smaller when aggregated over the relations among categories.

In summary, our results suggest that the robustness of the maps is due to a "pixelling effect": one can decompose a picture into different rather coarse subsets, and yet keep an understanding of the original image if it is watched (i.e., "integrated") at a distance. The overall quality of a map at a high level of aggregation does not inform us about the quality of the underlying micro-relations other than statistically. The "consensus" among these maps can thus be considered as an effect of the aggregation.

As a second important observation that can be made on the basis of these maps, we wish to point to the differences in category density between the content-based and the algorithm-based maps. In the case of the ISI and SOOI classifications (Figures 3 and 5, respectively), the three areas around biomedical sciences, chemistry, and physics/materials sciences contain a very high density of categories. The cosine similarity coefficients among these categories explain why: in the biomedical-science areas, the group of journals classified by the ISI as *Biochemistry and Molecular Biology* has very high similarity values with those in the category *Biophysics* (0.97), *Cell Biology* (0.95) and *Genetics and Heredity* (0.81). In the SOOI classification the category of *Biochemistry-Biophysics-Molecular Biology* has a cosine of 0.93 with *Cell Biology* and 0.74 with *Genetics and Developmental Biology*. However, in the Random Walk and Unfolding classifications, all these categories are brought together under a single large *Biomedical Sciences* category. The highest cosine similarity values within these two decompositions are only 0.67 and 0.58, respectively.

This example is meant to illustrate how the algorithmic decompositions subsume under a single category a variety of categories of the ISI and SOOI classifications. This similarity among content-based categories is particularly relevant for the assessment of "interdisciplinarity" or more generally "diversity" (Morillo *et al*., 2003; Stirling, 2007; Porter *et al.,* 2007; Rafols and Meyer, forthcoming). Using the ISI or SOOI classifications, a collection of papers published in the biosciences could be counted as interdisciplinary because the set is retrievable under a number of categories in the biomedical sciences (e.g., Cell Biology, Biophysics, and Developmental Biology).[11] However, if one takes into account that these categories have large overlaps among them (or, according to the cosine measure, are very similar), one reaches a different conclusion: this same document set can rather be considered as monodisciplinary within a single large disciplinary category as obtained by using classifications based on a Random Walk or the Unfolding algorithm.

---

[11] Most common measures of interdisciplinarity are based on numbers or the balancing of categories (e.g., Van Raan and Van Leewen, 2002) or indicators that subsume both number and balance (such as the Shannon entropy, e.g., Grupp, 1990; Rafols & Meyer, forthcoming).

The example discussed above suggests that in order to mitigate the biases introduced by classifications on analysis, classifications should be provided together with their similarity matrix and size distribution (as discussed in Tables 1 and 2). Although classifications can differ substantially, the similarity matrix and size distribution allow the user to contextualize the extent to which categories have overlaps, fuzzy areas or are very distinct. This can then be used to generate indicators of interdisciplinarity that take into account the degree of overlap between categories (Porter et *al.*, 2007; Rafols and Meyer, forthcoming).

In summary, the different science maps are surprisingly similar except that they differ in the density of categories within groups. The maps show in an intuitive way that the content-based classifications achieve a more balanced coverage of the disciplines at the expense of distinguishing categories that may be highly similar in terms of journals. These categories could have been collapsed without losing analytical quality. We appreciate that the ISI Subject Categories were primarily designed for information retrieval and not for scientometric purposes. However, the SOOI categories, which were developed for the purpose of scientometric evaluations, are even more densely packed than the ISI Subject Categories.

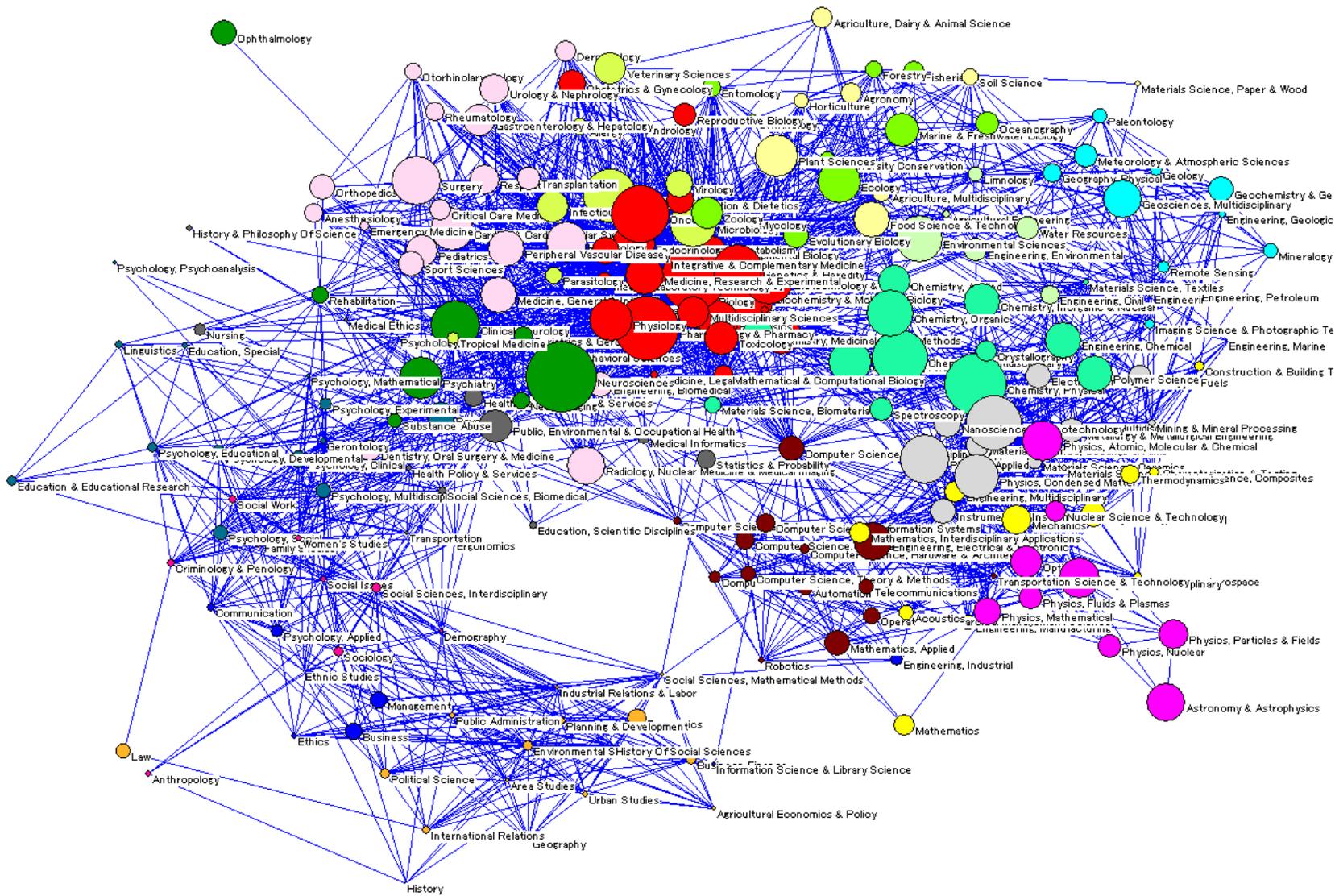

**Figure 3**. Similarity map among the 220 Subject Categories of the ISI; cosine > 0.2.

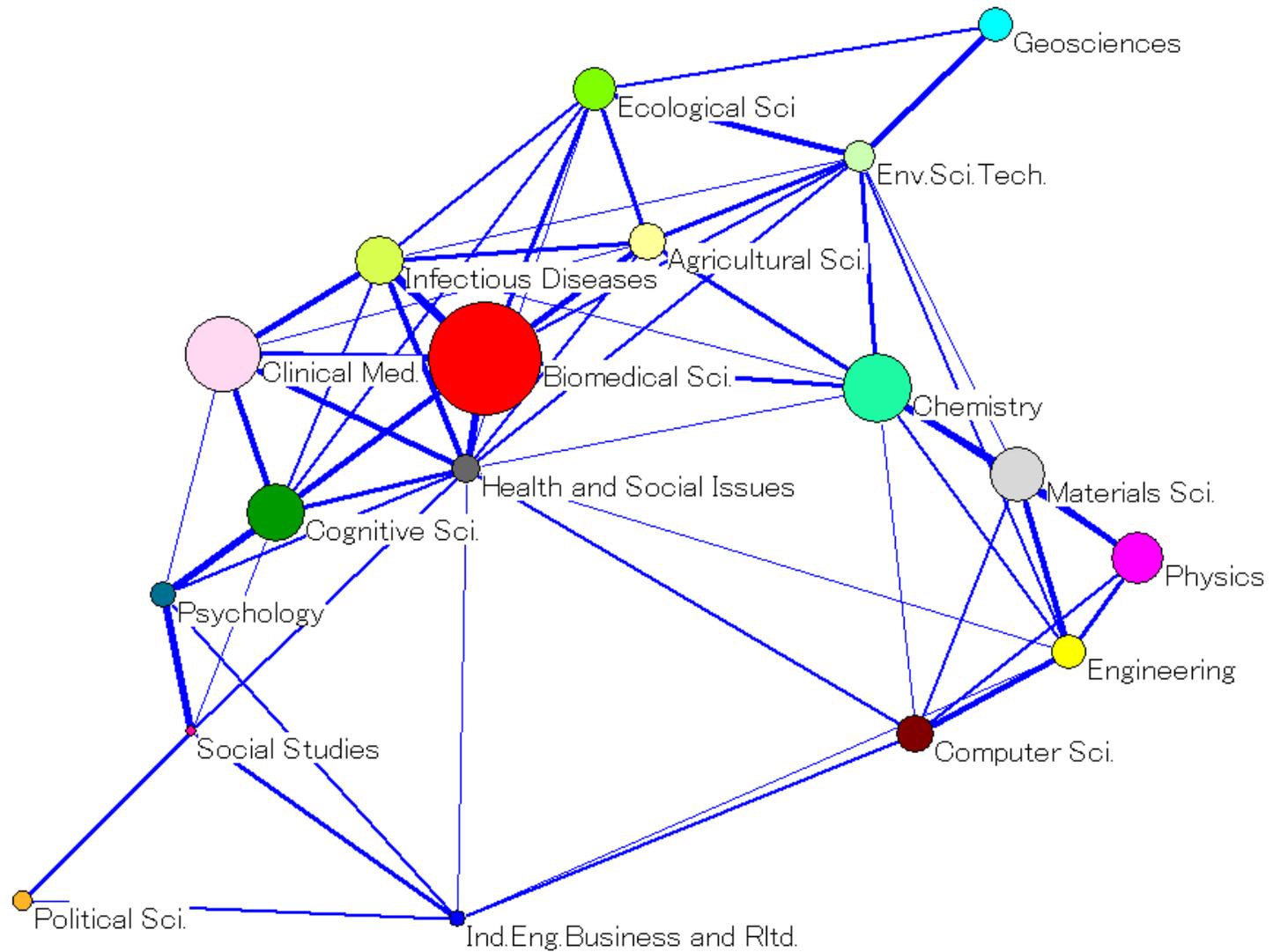

**Figure 4**. Similarity map among the 18 macro-categories on the basis of factor analysis of the 220 ISI Subject Categories; cosine > 0.1.

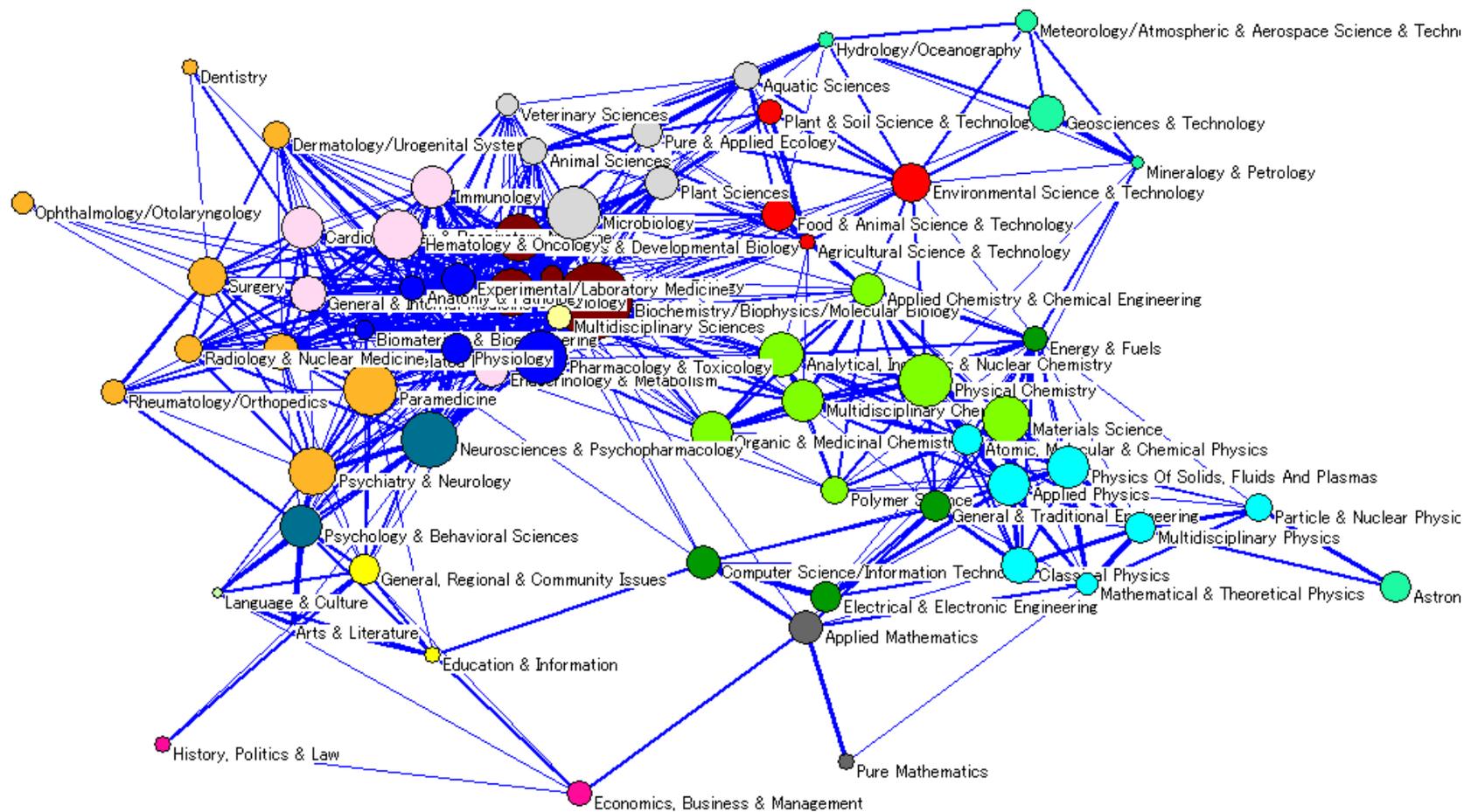

**Figure 5**. Similarity map among the 67 categories of the SOOI; cosine > 0.2.

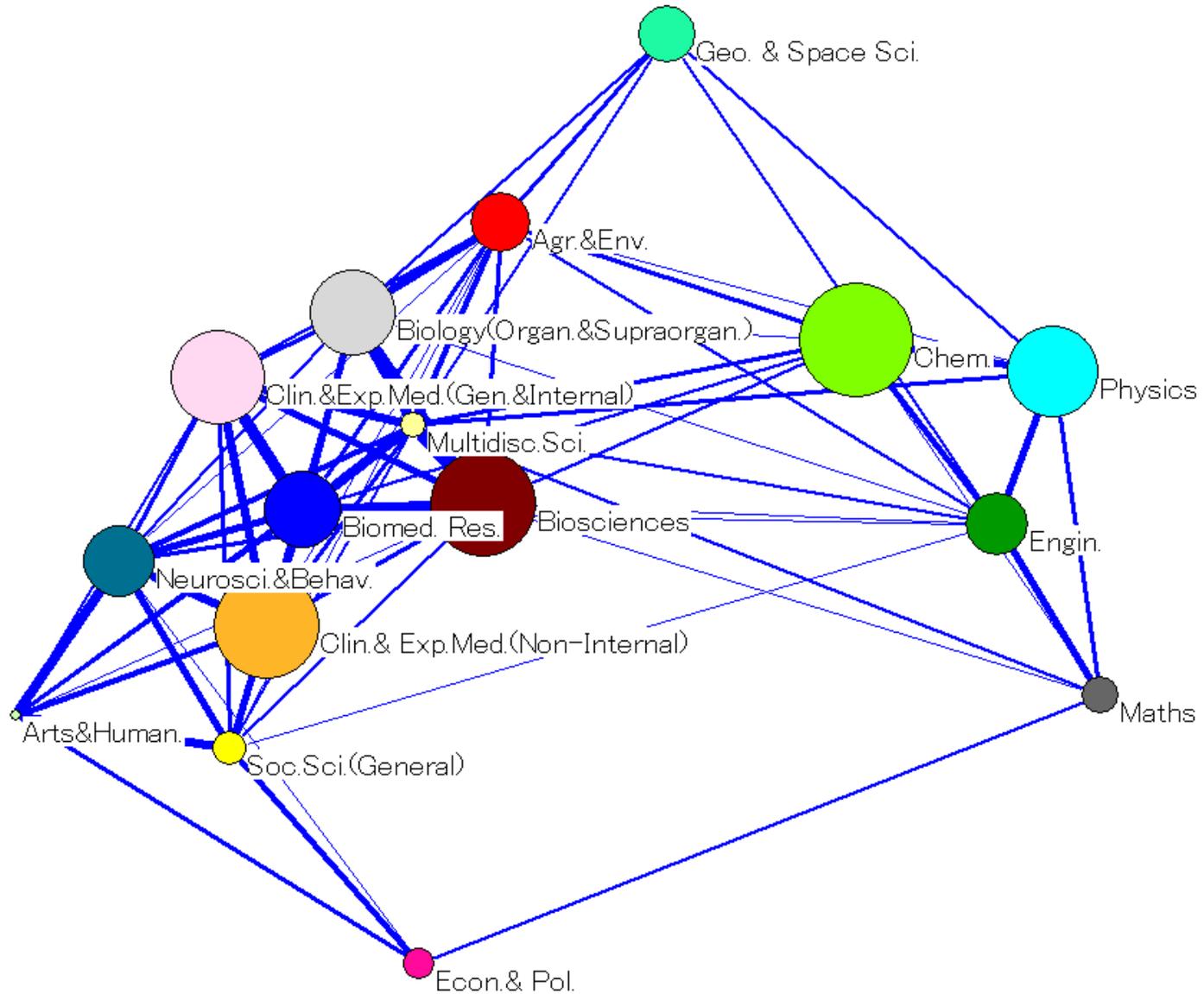

**Figure 6**. Similarity map among the 14 field-categories of the SOOI decomposition; cosine > 0.1.

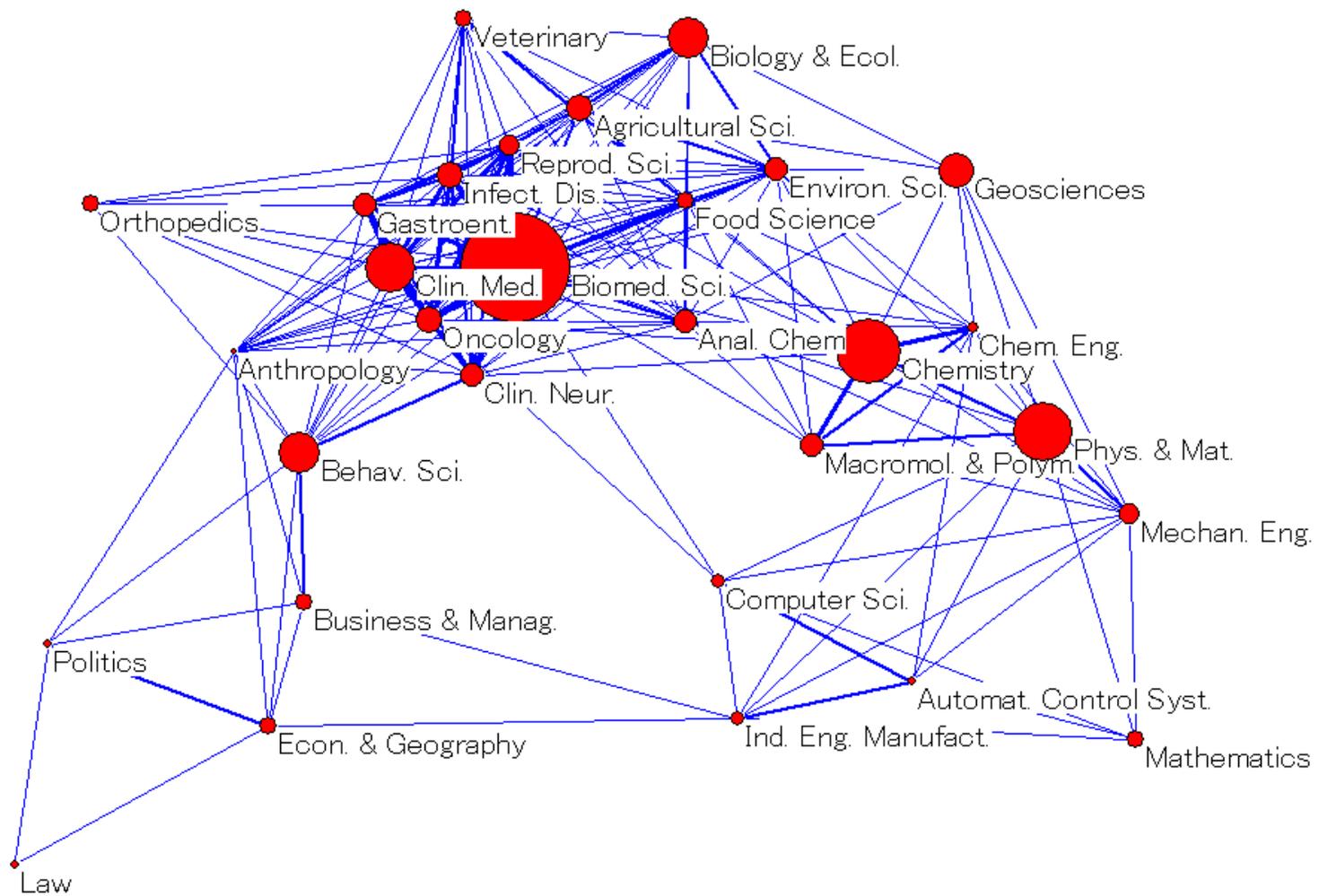

**Figure 7**. Similarity map among the first 30 categories of the decomposition based on a Random Walk algorithm; cosine > 0.02.

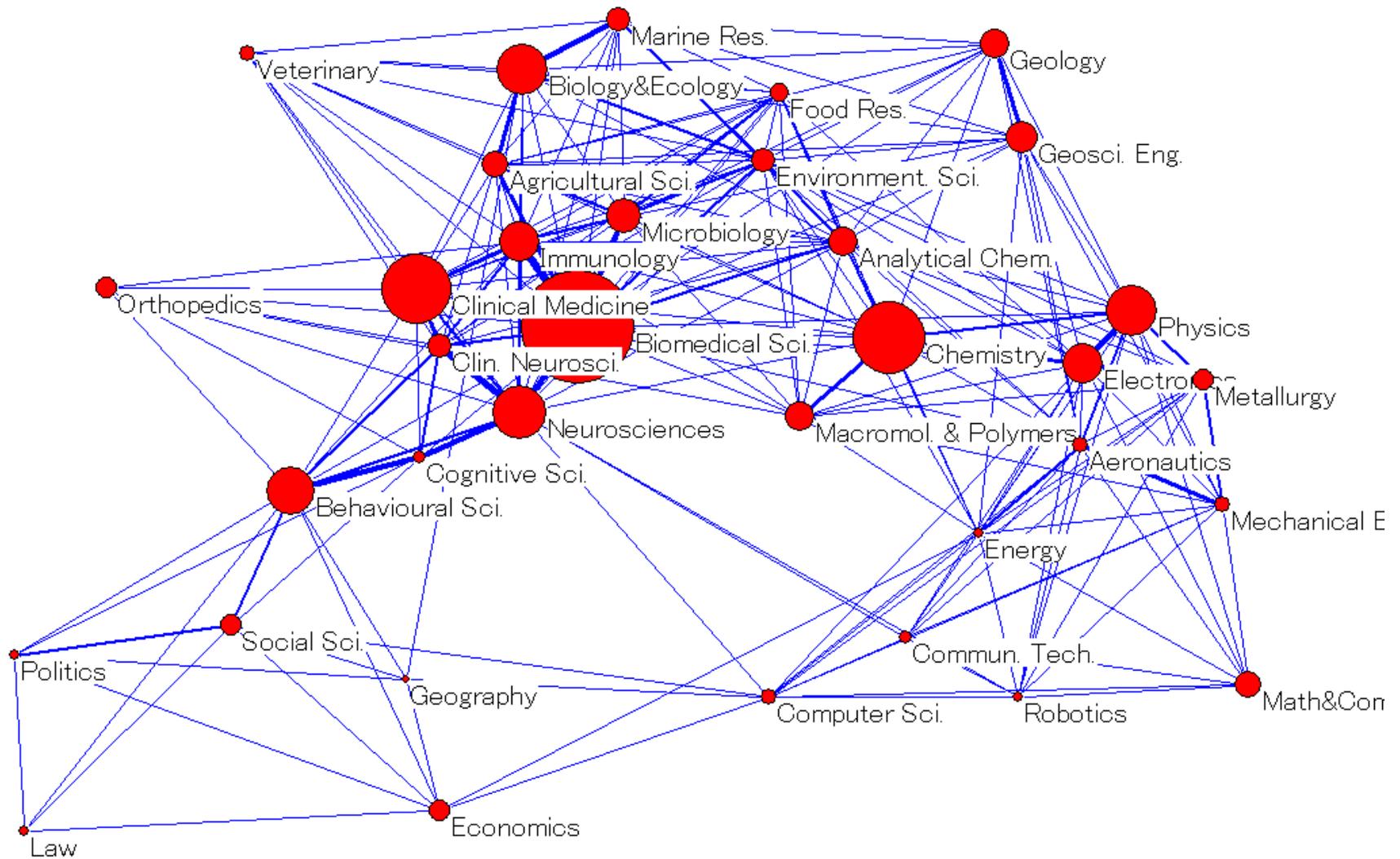

**Figure 8**. Similarity map among the first 35 categories of the decomposition based on the Unfolding algorithm; cosine > 0.02.

**Conclusions**

In this paper we have analyzed the results of four different decompositions of the aggregated journal-journal citation matrix. The first finding is that the algorithmic decompositions have very skewed and clean-cut distributions, with large clusters in a few scientific areas (e.g., Biosciences or Clinical Medicine), whereas indexers maintain more even and overlapping distributions in the content-based classifications. Second, the different classifications show a limited degree of agreement in terms of matching categories.[1] In spite of this lack of agreement, however, the science maps obtained are surprisingly similar; this robustness is due to the fact that although categories do not match precisely, their relative positions in the network among the other categories is based on distributions which match sufficiently to produce corresponding maps at the aggregated level.

Regarding the first finding, we suggest that because of the specific shape of the citation distributions among journals—with dense citation traffic within clusters and weak citation traffic among them—the algorithmic decompositions can be highly successful in decomposing the ISI journal set using maximization of an objective function. One would expect the results of the two algorithmic decompositions to be very similar because of the common ideas behind them (Newman, 2006a and b). Because of the log-normal shape of the resulting distribution, the number of clusters distinguished by each algorithm can be expected to determine the skewness (Figure 1). In other words, the results can be compared with hierarchical clustering when one changes the cut-off level: the lower the level, the more groups are distinguished, and larger groups can be expected to split up first when lowering the threshold level.

---

[1] These results are in accordance with previous findings comparing other classifications, as illustrated in Figure 12 in Leydesdorff (2006, p. 612).

At the other extreme, we found the ISI Subject Categories, which are mainly content-based. The SOOI categories mitigate the effects of the algorithmic approaches by adding content-based information to the citation-based information, and can thus be expected to occupy an intermediate position. However, the differences between the ISI Subject Categories and the SOOI field/subfield classifications in terms of these statistics were marginal. The largest categories in the ISI classification subsume 200+ journals, while the largest categories in the SOOI classification subsume 400+ journals. These differences are again an effect of the granularity of the classification: the ISI distinguishes 220 categories and the SOOI 67 subfields.

**Discussion**

Two strategies are available to the content-based indexer that cannot so easily be left to the machine algorithm. First, the indexer can be expected to be aware that categories should not be unbalanced in terms of size. Thus, there will be a tendency to split up large categories if possible, and to merge extremely small ones. This human intervention can be considered as a legitimate indexer effect. After all, indexing aims at facilitating the user to retrieve relevant information, and neither extremely large nor extremely small categories are functional to this end.

Secondly, the indexer can add cross-connections between categories which cannot be legitimated in terms of the nearly decomposable citation matrix. Both the ISI and the SOOI classifications make use of this option and thus generate off-diagonal elements. However, one should be aware that the newly introduced off-diagonal elements cannot be legitimated by the citation patterns generated by the scientists themselves, but are added by indexers.

Of course, indexers may consult individual experts—Glänzel & Schubert (2003) iteratively involve experts in each round of generating the classifications—but the communities of experts themselves neither produce nor reproduce these externally identified cross-connections. The cross-connections may provide a help function to the user of the database by assuming an external user's perspective to the underlying structures. Although these classifications are content-based, they no longer provide a systematic representation of the dynamics of scientific communication.

As noted, the content-based categorizations can be used for the mapping at the aggregate level because these maps use information regarding the relative proximity of categories. This suggests that the providers of these categorizations could add information regarding the size distribution of the categories and the similarities among them (underlying the mapping) in order to control for the indexer effects. However, for the purposes of more finer-grained evaluations, the indexer effects can be expected to disturb maps at lower levels of aggregation; for example, for the indication of interdisciplinarity.

In our opinion, the sociology of science—and along with it the policy analysis of science—should remain as close as possible to the representations of science that are reproduced in terms of aggregated citations by the scientists themselves, rather than those that are based on a reflexive representation of these data by indexers (or bibliometricians), who operate from different premises. However puzzling the results of the algorithmically generated classifications may be, these structures provide us with the *explanandum* of a theory of citations or a sociology of science. The reduction of uncertainty in hierarchical classifications by indexers cannot be taken for granted by these sciences. In other words, we do not wish to deny the pragmatic usefulness of the content-based categories, but their validity.

**Acknowledgements**

We are grateful to the following colleagues for providing data and assistance: Wolfgang Glänzel provided us with the attribution of the SOOI categories to the journals; Martin Rosvall guided us in using his algorithm for the random walk; Renaud Lambiotte offered to run the data using the algorithm for the Unfolding decomposition, and provided us with his results.

**Appendix A:** Example of the cross-tables between classifications.

For the purpose of illustrating the process of matching categories, Table A1 shows the cross-table between the seven largest categories in a Random Walk versus the 220 ISI Subject Categories. This table can be used to identify those categories that do not match between classifications and which therefore may be of policy interest. The mismatches may be indexer effects or reflect an interdisciplinary process. By analyzing the off-diagonal elements of the matrices, one can learn to what extent the delineations are fuzzy and which categories can be considered as complementary disciplinary sources for the fuzzy sets.

For example, 199 journals are subsumed under the ISI subject category of *Pharmacology and Pharmacy* of which only 89 are in the Biomedical Sciences according to the Random Walk, while 12 are organized in the Behavioural Sciences, and 40 in Clinical Medicine. At the other extreme, we find 100% agreement between the ISI Subject Categories and the Random Walk in the case of 36 *Palenteology* journals. This suggests that this is a very clearly defined research field. As noted, the overlap in the case of the Biomedical Sciences may be a spurious effect from the overlap among the ISI Subject Categories.

| Rows → Random Walk Category  ←Columns  ISI Subject Categories | Biomed. Sci. | Behav. Sci. | Clinical Med. | Biology & Ecology | Phys. & Materials | Geosci. | Econ. & Geogr. | Total Journals | % in top categ. |
|---|---|---|---|---|---|---|---|---|---|
| Biochem. Mol. Biol. | 206 | 1 | 5 | 8 | 0 | 0 | 0 | 262 | 79% |
| Cell Biology | 140 | 0 | 1 | 1 | 0 | 0 | 0 | 156 | 90% |
| Neurosciences | 115 | 38 | 0 | 1 | 0 | 0 | 0 | 199 | 58% |
| Genetics & Heredity | 97 | 4 | 3 | 13 | 0 | 0 | 0 | 131 | 74% |
| Pharmacol. & Pharm. | 89 | 12 | 40 | 0 | 0 | 0 | 0 | 199 | 45% |
| Biotech. & Microbiology | 86 | 0 | 4 | 5 | 0 | 0 | 0 | 140 | 61% |
| Immunology | 83 | 0 | 0 | 0 | 0 | 0 | 0 | 117 | 71% |
| Microbiology | 54 | 1 | 0 | 4 | 0 | 0 | 0 | 88 | 61% |
| Physiology | 50 | 3 | 2 | 6 | 0 | 1 | 0 | 79 | 63% |
| Med., Res. & Exp. | 49 | 0 | 17 | 0 | 0 | 0 | 0 | 76 | 64% |
| Endocrin. & Metabolism | 46 | 1 | 21 | 0 | 0 | 0 | 0 | 93 | 49% |
| Biophysics | 46 | 1 | 0 | 0 | 0 | 1 | 0 | 66 | 70% |
| Oncology | 44 | 0 | 1 | 0 | 0 | 0 | 0 | 127 | 57% |
| Toxicology | 43 | 0 | 5 | 0 | 0 | 0 | 0 | 76 | 57% |
| Biochem. Res. Methods | 35 | 0 | 0 | 0 | 1 | 0 | 0 | 56 | 63% |
| Biology | 29 | 0 | 0 | 16 | 0 | 2 | 0 | 64 | 45% |
| Hematology | 25 | 0 | 16 | 0 | 0 | 0 | 0 | 61 | 41% |
| Developmental Biology | 25 | 0 | 0 | 0 | 0 | 0 | 0 | 34 | 74% |
| Psychiatry | 3 | 118 | 0 | 0 | 0 | 0 | 0 | 133 | 89% |
| Psych., Multidisciplinary | 0 | 80 | 0 | 0 | 0 | 0 | 1 | 84 | 95% |
| Education | 0 | 74 | 0 | 0 | 0 | 0 | 3 | 93 | 80% |
| Psychology, Clinical | 0 | 68 | 0 | 0 | 0 | 0 | 0 | 70 | 97% |
| Psychology, Developmental | 0 | 47 | 0 | 0 | 0 | 0 | 0 | 47 | 100% |
| Psychology | 2 | 46 | 0 | 0 | 0 | 0 | 0 | 61 | 75% |
| Rehabilitation | 0 | 45 | 1 | 0 | 0 | 0 | 0 | 72 | 63% |
| Psychology, Social | 0 | 42 | 0 | 0 | 0 | 0 | 0 | 46 | 91% |
| Psychology, Experimental | 0 | 42 | 0 | 0 | 0 | 0 | 0 | 44 | 95% |
| Psychology, Educational | 0 | 38 | 0 | 0 | 0 | 0 | 0 | 38 | 100% |
| Medicine, General & Internal | 6 | 2 | 86 | 0 | 0 | 0 | 0 | 103 | 83% |
| Cardiovascular Systems | 5 | 0 | 65 | 0 | 0 | 0 | 0 | 74 | 88% |
| Public, Env.& Occ. Health | 3 | 32 | 52 | 0 | 0 | 0 | 0 | 141 | 37% |
| Health Care Sci. | 0 | 2 | 46 | 0 | 0 | 0 | 0 | 56 | 82% |
| Periph. Vascular Disease | 9 | 0 | 39 | 0 | 0 | 0 | 0 | 52 | 75% |
| Pediatrics | 2 | 5 | 32 | 0 | 0 | 0 | 0 | 74 | 43% |
| Nutrition & Dietetics | 8 | 1 | 30 | 0 | 0 | 0 | 1 | 55 | 55% |
| Surgery | 2 | 0 | 12 | 0 | 0 | 0 | 0 | 138 | 28% |
| Med. Lab. Technology | 7 | 0 | 11 | 0 | 0 | 0 | 0 | 25 | 44% |
| Geriatrics & Gerontology | 8 | 6 | 10 | 0 | 0 | 0 | 0 | 30 | 33% |
| Medical Informatics | 1 | 0 | 10 | 0 | 0 | 0 | 0 | 20 | 50% |
| Ecology | 0 | 0 | 0 | 104 | 0 | 2 | 1 | 114 | 91% |
| Zoology | 9 | 0 | 0 | 85 | 0 | 0 | 0 | 114 | 75% |
| Marine Biology | 1 | 0 | 1 | 73 | 0 | 0 | 0 | 79 | 92% |
| Entomology | 5 | 0 | 0 | 58 | 0 | 0 | 0 | 69 | 84% |
| Mat. Sci., Multidiscipl. | 0 | 0 | 0 | 0 | 94 | 1 | 0 | 175 | 54% |
| Physics, Applied | 0 | 0 | 0 | 0 | 69 | 1 | 0 | 84 | 82% |
| Metal. & Metal. Eng. | 0 | 0 | 0 | 0 | 51 | 0 | 0 | 65 | 78% |
| Geosciences, Multidiscipl. | 1 | 0 | 0 | 4 | 0 | 101 | 0 | 131 | 77% |
| Geochem. & Geophysics | 0 | 0 | 0 | 0 | 0 | 58 | 0 | 59 | 98% |
| Meteorology | 0 | 0 | 0 | 0 | 0 | 46 | 0 | 48 | 96% |
| Paleontology | 0 | 0 | 0 | 0 | 0 | 36 | 0 | 36 | 100% |
| Geology | 0 | 0 | 0 | 0 | 0 | 35 | 0 | 36 | 97% |
| Oceanography | 0 | 0 | 0 | 18 | 0 | 24 | 0 | 48 | 50% |
| Mineralogy | 0 | 0 | 0 | 0 | 1 | 23 | 0 | 26 | 88% |
| Geography, Physical | 0 | 0 | 0 | 7 | 0 | 22 | 0 | 30 | 73% |
| Economics | 0 | 0 | 0 | 0 | 0 | 0 | 145 | 154 | 94% |
| Business, Finance | 0 | 0 | 0 | 0 | 0 | 0 | 37 | 39 | 95% |
| Geography | 0 | 0 | 0 | 0 | 0 | 0 | 36 | 36 | 100% |
| Environmental Studies | 0 | 4 | 0 | 0 | 0 | 0 | 35 | 45 | 78% |
| Planning & Development | 0 | 1 | 0 | 0 | 0 | 0 | 29 | 37 | 78% |
| Urban Studies | 0 | 2 | 0 | 0 | 0 | 0 | 23 | 27 | 85% |
| Total # of journals in category | 942 | 722 | 456 | 437 | 375 | 301 | 301 | | |
| % in top category | 22% | 16% | 19% | 24% | 25% | 34% | 48% | | |

**Table A1.** Number of shared journals between categories of the Random Walk decomposition (columns) and the ISI Subject Categories (rows).